\documentclass[aps,twocolumn,showpacs,amsmath,amssymb,pre,superscriptaddress, floatfix]{revtex4-1}

\usepackage{graphicx}
\usepackage{dcolumn}
\usepackage{bm}
\usepackage{amssymb}
\usepackage{multirow}
\usepackage{braket}

\begin{document}

\title{Site-selective particle deposition in periodically driven quantum lattices}

\author{Thomas Wulf}
    \email{Thomas.Wulf@physnet.uni-hamburg.de}
    \affiliation{Zentrum f\"ur Optische Quantentechnologien, Universit\"at Hamburg, Luruper Chaussee 149, 22761 Hamburg, Germany}

\author{Benno Liebchen}
    \affiliation{SUPA, School of Physics and Astronomy, University of Edinburgh, Edinburgh EH9 3FD, United Kingdom}

\author{Peter Schmelcher}
    \email{Peter.Schmelcher@physnet.uni-hamburg.de}
    \affiliation{Zentrum f\"ur Optische Quantentechnologien, Universit\"at Hamburg, Luruper Chaussee 149, 22761 Hamburg, Germany}
    \affiliation{The Hamburg Centre for Ultrafast Imaging, Universit\"at Hamburg, Luruper Chaussee 149, 22761 Hamburg, Germany} 

\date{\today}


\begin{abstract}

We demonstrate that a site-dependent driving of a periodic potential allows for 
the controlled manipulation of a quantum particle on length scales of the lattice spacing.
Specifically we observe for distinct driving frequencies a near depletion of certain sites which is explained by a resonant mixing of the involved
Floquet-Bloch modes occurring at these frequencies. Our results could be exploited as a scheme for 
a site-selective loading of e.g. ultracold atoms into an optical lattices.

\end{abstract}

\maketitle

\paragraph*{Introduction}

The nonequilibrium physics of driven lattices has long been the subject of intensive theoretical research 
\cite{Hanggi:2005, Hanggi:2009, Arimundo:2012, Hanggi:2014} with applications
to numerous experimental setups such as terahertz driven semiconductor heterostructures \cite{Olbrich:2009, Olbrich:2011}
or cold and ultracold atoms loaded into driven optical lattices \cite{Renzoni:2010, Salger:2009, Weitz:2013}. 
Thereby, it was shown that the interplay of a spatially periodic lattice potential and a driving force leads to a plethora 
of interesting non equilibrium phenomena, a paradigmatic 
example being the celebrated 'ratchet effect'  where
particles undergo directed motion despite the absence of any mean forces \cite{Flach:2000, Renzoni:2012, Schanz:2005, Salger:2009}. 
Besides that and triggered particularly by the upcoming ultracold atom experiments the inclusion of a driving force was used for 
the renormalization of the tunneling rates between adjacent lattice sites \cite{Lignier:2007} or for the engineering of so called artificial 
gauge fields \cite{Struck:2012, Goldman:2014}. \\
While the main focus has so far been
on global driving forces that are the same everywhere in space, 
it has recently been shown how local modulations 
of the driving allow for extensive manipulations of the particles classical dynamics leading to phenomena such as a site-dependent particle trapping \cite{Petri:2010, Liebchen:2011}, 
the spontaneous formation of density waves or the emergence of order by the combination of disorder and driving \cite{Petri:2011, Wulf:2012, Wulf:2014}.
Only very recently, 
this concept of a spatially varying driving has
been applied firstly in the quantum domain \cite{Wulf:2014_2}, where the transformation of an 
avoided- to an exact crossing in the Floquet spectrum 
as well as the control of asymptotic currents have been shown
and explained as a consequence of the local driving.
In the present work we demonstrate how the site-dependent driving opens 
the roadway towards a control of the nonequilibrium dynamics of wave packets on length scales of the lattice spacing.
We propose how, by exploiting a resonant mixing of Floquet-Bloch modes, quantum states which are highly localized on specific lattice sites can be prepared in a very controlled manner.
The in this way achieved spatially dependent occupation of the lattice sites goes clearly beyond previously introduced schemes such as 'light shift engineering' \cite{Griffin:2006} 
where a spatial dependence of the atom density was obtained only on length scales of several hundred lattice sites.

\begin{figure}[htbp]
\centering
\includegraphics[width=1\columnwidth]{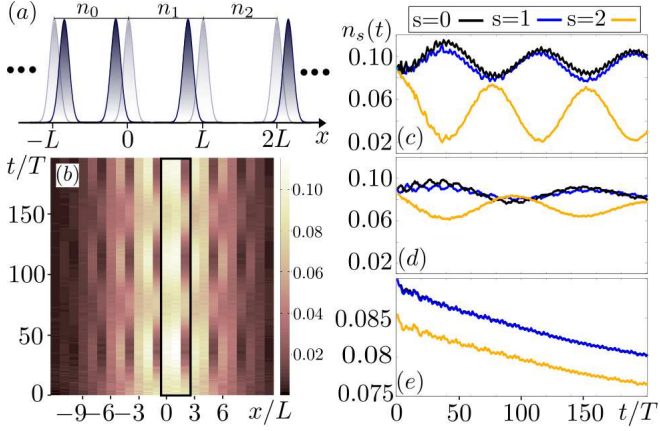}
\caption{\label{fig1} (a) Snapshot of a site-dependently driven lattice with lattice site occupations $n_s$. Shaded barriers indicate the barriers equidistant equilibrium positions.
(b) Time evolution of lattice site populations for an initial Gaussian of width $\sigma=20\pi$ in a lattice with 3 barriers per unit cell with phases $(0,\pi,0)$.
(c), (d) and (e) depict populations of the first three lattice sites (encompassed by a black rectangle in (b)) for phases of the central barrier in each unit cell of $\pi$ (c), $0.2\pi$ (d) and $0$ (e).
Remaining parameters are $L=10, \ V_0=1.0, \ \omega=1.0, A=1.0, m=1.0, \hbar=1.0$ and $\Delta=0.5$.}
\end{figure}

\paragraph*{Setup}
\label{S1}
We consider quantum particles in a time-dependent periodic potential of laterally oscillating Gaussian barriers in one dimension (Fig. \ref{fig1} (a)).
The dynamics is governed by the time-dependent Schr\"{o}dinger equation (TDSE) $i \hbar \frac{\partial}{\partial t}\Psi(x,t) =H(x,t) \Psi(x,t)$
with the Hamiltonian:
\begin{equation}
  H(x,t)=-\frac{\hbar^2}{2m}  \frac{\partial^2}{\partial x^2} + V_0 \sum_{s=-\infty}^{\infty} e^{-\left(  \frac{x-s\, L- d_s(t)}{\Delta}      \right)^2 }.
\label{Hamiltonian}
\end{equation}
with $m, V_0, \Delta$ and $L$ being the particle mass, potential height, barrier width and lattice spacing respectively.
$d_s(t)=A\cos(\omega t + \delta_s)$ is the driving law of site $s$ with frequency $\omega$ and amplitude $A$.
We will restrict ourselves to cases with 
$\delta_s=\delta_{s+n_p}$ for some $n_p \in \mathbb{N}$ such that $n_p L$ is the spatial 
period of the lattice.

\paragraph*{Formalism and computational scheme}
\label{S2}
We briefly outline the employed computational method \cite{Wulf:2014_2}.
The temporal periodicity of the Hamiltonian [$H(x,t)=H(x,t+T)$ with $T=2\pi/\omega$]
ensures that every solution of the TDSE can be written as
$\Psi_{\alpha}(x,t)=e^{-i\epsilon_{\alpha}t/\hbar} \Phi_{\alpha}(x,t)$ with 
the real quasi energy (QE) $\epsilon_{\alpha} \in [-\hbar\omega/2, +\hbar\omega/2]$ and the Floquet mode $\Phi_{\alpha}(x,t)$ obeying 
$\Phi_{\alpha}(x,t) = \Phi_{\alpha}(x,t+T)$. 
Due to the additional spatial periodicity [$H(x,t)=H(x+n_p L,t)$]
each Floquet mode can be written in terms of a Floquet-Bloch mode (FBM) as 
$\Phi_{\alpha, \kappa}(x,t)=e^{i\kappa x} \phi_{\alpha, \kappa}(x,t)$ with  
$\phi_{\alpha, \kappa}(x,t)= \phi_{\alpha, \kappa}(x+ n_p L,t)$ and $\kappa \in [-\pi/(n_p L),+\pi/(n_p L)]$ 
being the quasi-momentum. 
For every value of $\kappa$, the FBMs are eigenstates of the 
time evolution operator over an entire period of the driving $U^{\kappa}(T+t_0,t_0)$ \cite{Tannor}:
\begin{equation}
 U^{\kappa}(T+t_0,t_0)\Phi_{\alpha, \kappa}(x,t_0)=e^{-i\epsilon_{\alpha, \kappa} T/\hbar}\Phi_{\alpha, \kappa}(x,t_0).
 \label{eigenvectors}
\end{equation}
Hence, by using that the FBMs constitute an orthonormal basis of the Hilbert space spanned by the solutions of the TDSE,
the stroboscopic time evolution of any initial state $\Psi(x,t_0)$ is given as:
\begin{equation}
\begin{aligned}
 \Psi(x,t_0+mT) &= \\
 \int_{-\pi / (n_p L)}^{+\pi / (n_p L)} &d\kappa \sum_{\alpha} C_{\alpha,\kappa}(t_0) e^{-i\epsilon_{\alpha,\kappa}mT/\hbar } \Phi_{\alpha,\kappa}(x,t_0).
  \label{time}
\end{aligned}
\end{equation}
where $C_{\alpha,\kappa}(t_0)$ are the overlap integrals between the initial state $\Psi(x,t_0)$ and the FBM $\Phi_{\alpha,\kappa}(x,t_0)$. 
After numerically diagonalizing the time evolution 
operator and thus obtaining the FBMs as its eigenstates we are able to propagate arbitrary initial states by employing Eq. (\ref{time}). 
Note that in principle, the FBMs and thus the dynamics of an initial state depend on the initial time $t_0$. However, the findings of this work
are very robust with respect to variations of $t_0$ and thus we will only present results for the case $t_0=0$ in the following.

\paragraph*{Lattice site occupations for site-dependent driving}
\label{S3}

We now demonstrate how a site-dependent driving can be used to modulate and control the occupation 
of specific lattice sites. 
We consider a Gaussian initial state $\Psi(x,0)=(\pi \sigma^2)^{-1/4} e^{-\frac{x^2}{2\sigma^2}}$
that is exposed to a lattice with three barriers per unit cell ($n_p=3$) and $\delta_s=(0,\delta,0)$ for $s=0,1,2$ and with $\delta \in [0,\pi]$.
The occupation of lattice site $s$ after $m$ periods of the driving is given by:
\begin{equation}
n_s(mT)=\int_{(s-1)L}^{sL} |\Psi(x,mT)|^2 \, dx 
\end{equation}
As a striking feature we observe pronounced and site-dependent temporal oscillations of the population $n_s(t)$ with an approximate period of $75T$ (Fig.~\ref{fig1}(b)).
Fig.~\ref{fig1}(c) reveals an additional, seemingly irregular, 
micro oscillation of the population $n_s(mT)$.
In addition, a slow overall decay of all three in Fig.~\ref{fig1}(c) depicted populations is observed, which is a straightforward consequence of the diffusion of the initial state. 
Note that the observed oscillations of the population crucially 
rely on the site-dependent driving and disappear when the phase shift $\delta$ of the central barrier approaches zero (Figs.~\ref{fig1}(d) and (e)).
In the following we will analyze and explain the oscillatory behaviour of the site populations $n_s(mT)$ and demonstrate afterwards how it can be exploited for
a selective loading of the lattice.\\
\begin{figure}[htbp]
\centering
\includegraphics[width=1\columnwidth]{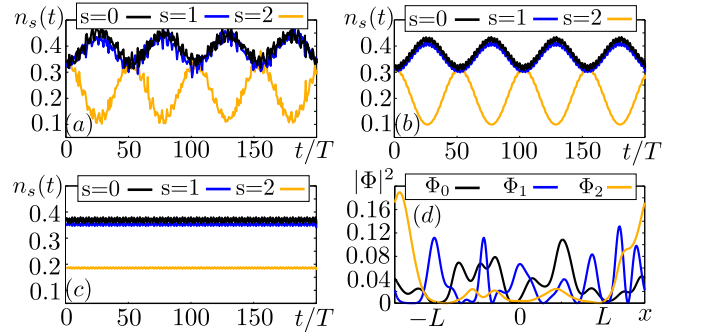}
\caption{\label{fig2} Setup consisting of three barriers with phases $\delta_s=(0,\pi,0)$ for periodic boundary conditions.
(a) Time evolution of the lattice site occupation $n_s(mT)$, (b) and (c) show the same quantity but only the 3 (b) or 2 (c) most occupied FBMs are taken into account, (d) position representation 
of the three most occupied FBMs. Remaining parameters as in Fig.~\ref{fig1}.}
\end{figure}
To understand the origin of the observed phenomena
let us systematically simplify the setup.
As a first step, we investigate the population dynamics $n_s(mT)$
in a setup consisting of only a single unit cell with periodic boundary conditions and with a uniform initial state 
(resembling the conditions at the center of a broad Gaussian initial state in an infinitely extended lattice).
Note that the inclusion of periodic boundary conditions corresponds to a 
restriction to $\kappa=0$, as every FBM fulfills 
$\Phi_{\alpha, \kappa}(x+n_pL,t)=e^{i n_p L \kappa}\Phi_{\alpha, \kappa}(x,t)$.
In fact Fig.~\ref{fig2}(a) reveals that our simplified lattice is capable of reproducing the key features of the population dynamics
observed in Fig~\ref{fig1}(c).
Further simplifying the setup, we now perform a few mode approximation 
after sorting the FBMs $\Phi_{\alpha}$ at $\kappa=0$ according to their overlap with the uniform initial state. 
Hence, the 'Floquet ground state' (FGS) corresponding to the largest overlap 
is labeled as $\Phi_{0}$, the mode with the second largest overlap as $\Phi_{1}$ etc. 
Evidently, for the three mode approximation (i.e. we set  $C_{\alpha}=0$ for $\alpha>2$ in Eq.~\ref{time}) the pronounced oscillation of the
lattice site population is reproduced (cf Fig.~\ref{fig2}(b)),
while it is not for the case of only two modes (cf Fig.~\ref{fig2}(c)). 
This suggests that the dynamics of the lattice site occupation is predominantly determined by the FBMs 
$\Phi_{0}$ and $\Phi_{2}$, as $\Phi_{1}$ does not contribute to the pronounced oscillations (cf Fig.~\ref{fig2}(c) and see discussion below).
Let us therefore assume that only these two modes are initially 
occupied, i.e. $\Psi(x,0)=C_0 \Phi_{0}(x,0) + C_2 \Phi_{2}(x,0)$. According to Eq.~\ref{time} 
we obtain: 
\begin{equation}
\begin{aligned}
 |\Psi(x,mT)|^2 &\approx |C_0 \Phi_{0}(x,0)|^2 + |C_2 \Phi_{2}(x,0)|^2 + \\
 &2 \text{Re}(C_0 C_2^*\Phi_{0}(x,0)\Phi_{2}^*(x,0) e^{im(\epsilon_2-\epsilon_0)T/\hbar}).
\end{aligned}
 \label{interference}
\end{equation}
The interference term oscillates with a period of $T_{\text{osc}}/T=\hbar \omega / (\epsilon_2-\epsilon_0)\approx 77$, which fits 
the period of the population oscillations
in the infinitely extended lattice (Fig.~\ref{fig1}(b)) quite well.
A remaining question is why a few mode approximation which includes $\Phi_{2}$ leads to pronounced oscillations of $n_s(mT)$, 
but not the corresponding approximation based solely on $\Phi_{0}$ and $\Phi_{1}$.
The reason for this is that the interference term in
Eq. \ref{interference} can only lead to a pronounced oscillation of $n_s(mT)$ if at least one of the involved FBMs is strongly localized on one of the lattice sites.
As Fig. \ref{fig2} (d) reveals, this is indeed the case for $\Phi_{2}$, but not for $\Phi_{1}$. Hence, even though the FBM $\Phi_{1}$ is more populated than $\Phi_{2}$,
it contributes much less to the distinct time evolution of the observable $n_s(mT)$.
\begin{figure}[htbp]
\centering
\includegraphics[width=1\columnwidth]{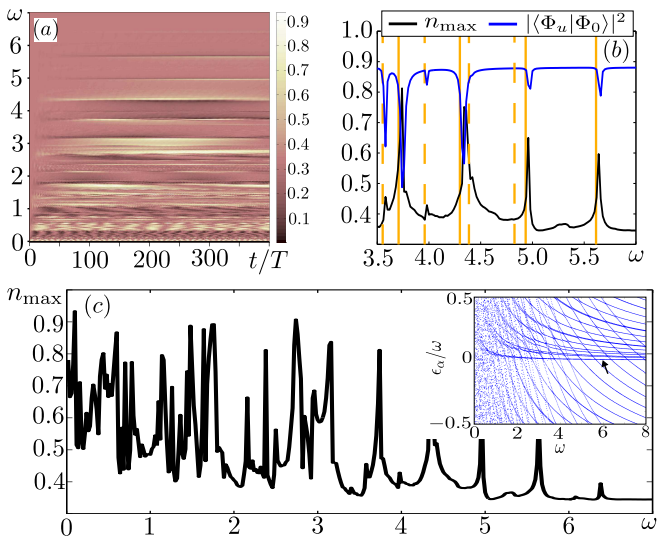}
\caption{\label{fig3}(a) Population of the $s=0$ lattice site for a setup containing three lattice sites with phases $(0,\pi,0)$ and periodic boundary conditions ($\kappa=0$) for different driving frequencies $\omega$.
(b) Extract of $n_{\text{max}}(\omega)$ (black) together with the overlap of the FGS with a uniform state $\Phi_u$ (blue). 
Vertical lines are by Eq. (\ref{resos}) predicted resonances for $n=1$ (solid) 
and $n=2$ (dashed).
(c) Maximal population of any of the 3 lattice sites $n_{\text{max}}(\omega)$ which is reached within the first 400 driving periods.  
Inset of (c) shows the QE spectrum 
(small arrow indicates the FGS). 
Remaining parameters as in Fig. \ref{fig1}.}
\end{figure}

\paragraph*{Controlling the lattice site populations}
\label{S5}
We now demonstrate the controllability of the time-dependent populations of the individual lattice sites via the driving frequency $\omega$. 
First, we keep the periodic boundary conditions, i.e.  $\kappa=0$, and
discuss the impact of nonzero $\kappa$ afterwards.
In the low frequency regime ($\omega \lesssim 3.5$), the population of site $s=0$, $n_0(mT)$,
exhibits oscillations whose period 
depends sensitively on $\omega$ thereby ranging from $T_{\text{osc}} \sim 5T$ to $T_{\text{osc}} > 400T$ (Fig.~\ref{fig3}(a)).
Contrarily, for $\omega \gtrsim 3.5$, $n_0(mT)$ becomes approximately constant with $n_0(mT) \approx 1/3$ 
except at very narrow frequencies intervals (e.g. at $\omega \approx 5$) where $n_0(mT)$ again strongly deviates from $1/3$. 
Since the population dynamics of the sites $s=1,2$ is qualitatively similar we refrain from showing it. 
Instead, we show the maximal occupation $n_{\text{max}}$ of any of the three lattice sites 
that is reached during the first $400$ driving periods as a function of $\omega$ (Fig.~\ref{fig3}(c)). 
The regime $\omega \lesssim 3.5$ exhibits an irregular behaviour reflecting the above observation of a high 
sensitivity of the lattice site populations on $\omega$.
Besides narrow peaks, we observe for $\omega \gtrsim 3.5$ that $n_{\text{max}}(\omega)$ is approximately $1/3$ corresponding to an 
equal population of the three sites at all times.
The overall convergence for $\omega \rightarrow \infty$ of $n_s(t)$ towards $1/3$ is not surprising, as for fast driving
the Hamiltonian can typically be approximated by its time average, which is site-independent in our case. 
Thus, it remains an intriguing question why the occupation of the different lattice sites inherits 
such a strong imbalance at distinct driving frequencies even in the regime of large $\omega$.
To gain insight, let us recall that the oscillations in the population dynamics $n_s(mT)$ 
can be attributed mostly to the interference 
of the FGS $\Phi_{0}$ with one of the higher modes. 
Since $\Phi_{0}$ is the dominantly occupied FBM, a strong 
desymmetrization of the lattice site populations as indicated by a large value of $n_{\text{max}}$ can be expected to correspond to a large desymmetrization of $\Phi_{0}$. 
Measuring the desymmetrization of $\Phi_{0}$ by its overlap with a uniform state $\Phi_u=1/\sqrt{n_p L}$,
we find that the peaks in $n_{\text{max}}(\omega)$ are indeed accompanied by resonant dips 
of the overlap  of $\Phi_{0}$ with $\Phi_u$ (Fig.~\ref{fig3}(b)). 
\\The remaining open question concerns the physical mechanism underlying the observed
desymmetrization of $\Phi_{0}$ at specific frequencies.
Here, the frequency-dependent QE spectrum reveals further insight (Fig. \ref{fig3}(c) inset). 
While the spectrum appears irregular at small $\omega \lesssim 2$ we see clearly the emergence of QE bands at larger $\omega$. 
Moreover, these bands are subject to multiple exact- and avoided
crossings. Of particular interest to our analysis is the FGS $\Phi_{0}$ (marked by the arrow in Fig.~\ref{fig3}(c) inset). 
A closer inspection reveals that each of
the peaks in $n_{\text{max}}(\omega)$ and equally the dips 
in $|\braket{\Phi_{0} | \Phi_u}|^2$
fall together with avoided crossings of $\Phi_{0}$ with states belonging to higher bands. Hence, the 
FGS mixes strongly with another state at these frequencies which leads to the strong desymmetrization and 
thus to a resonant enhancement of $n_{\text{max}}(\omega)$. 
By exploiting this insight we can employ a remarkably simple approximation in order to deduce the resonance
positions in frequency space.
As FBMs with average kinetic energies $\gg V_0$ represent an almost free propagation through the lattice 
we approximate $V_0 \approx 0$ and calculate their QEs easily 
as $\epsilon_{\alpha}=2 \pi^2 \alpha^2 /m(n_P L)^2$ (which then have to be folded back to the first Brillouin zone).
In this regime of fast driving we may further assume $\epsilon_0\approx 0$ compared to the extension of the Brillouin zone $\hbar \omega $. 
Consequently, the frequencies $\omega_{\text{res}}$ corresponding to resonances in $n_{\text{max}}(\omega)$ caused by avoided 
crossings of the FGS and the higher bands, 
can be approximated as the zeros of the free particle quasi energy:  
\begin{equation}
 \omega_{\text{res}}(\alpha,n)=\frac{2\pi^2 \hbar}{ m (n_p L)^2} \frac{\alpha^2}{n}, \quad \text{for} \ n=1,2,3,...
 \label{resos}
\end{equation}
Here, $n$ denotes the number of times the free particle energy has to be folded back into the first Brillouin zone. 
In Fig.~\ref{fig3}(b) we show the 
locations of the resonances predicted by Eq. (\ref{resos}) for $n=1,2$ and up to $\alpha=20$, corresponding to the number of FBMs which are taken into account numerically and which are shown in
the spectrum (cf. inset Fig. \ref{fig3} (c)),
confirming a good agreement. \\
 
\begin{figure}[htbp]
\centering
\includegraphics[width=1\columnwidth]{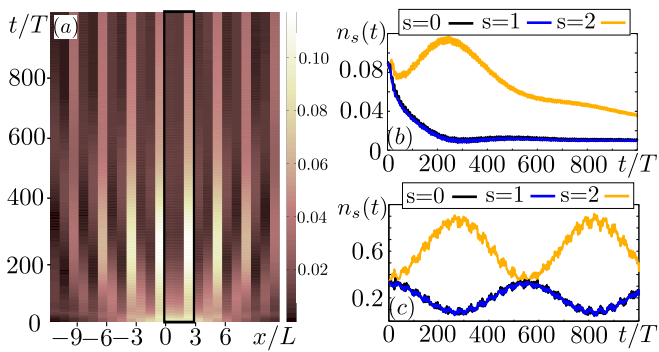}
\caption{\label{fig4} (a) Time evolution of the lattice site occupation $n_s(mT)$ for the resonant case $\omega=2.74$. 
(b) $n_s(mT)$ for $s=0,1,2$ (encompassed by a black rectangle in (a)). (c) same as in (b) but restricted to $\kappa=0$. The $s=0$ and $s=1$ curves in (b) and (c) are barely distinguishable.
Remaining parameters as in Fig. \ref{fig1}. }
\end{figure}
 
\paragraph*{Resonant population transfer at nonzero quasi-momentum}
\label{S6}
Finally, we demonstrate how the above analyzed resonant population imbalance is altered by the inclusion of higher quasi-momenta.  
We explore the time evolution of the same Gaussian wave packet as 
above (Fig. \ref{fig1}(b)) but now with a driving frequency $\omega=2.74$ where $n_{\text{max}}(\omega)$ features one of its strongest peaks.
After a transient of $t \sim 50 T$, we observe a pronounced imbalance of the lattice site populations reaching its maximum at $t/T\approx 250$ (Fig.\ref{fig4} (a) and (b)). 
Interestingly and in clear contrast to the $\kappa=0$ case (Fig.~\ref{fig4}(c)), we observe no indication of a pronounced oscillation of population between the sites. 
Apparently, the inclusion of higher quasi-momenta leads to a decay of the oscillation amplitudes. Thus, while we still observe an overall diffusive decay at large times, 
the imbalance between the lattice sites is effectively frozen at intermediate times ($500 \lesssim t/T \lesssim 800$).
Note that in the previously studied case of $\omega=1$ (Fig \ref{fig1} (c)) we could not see this effect as the oscillations there occur on much faster time scales compared 
to the decay of the oscillation amplitude.
To gain further insight, let us make some simplifying assumptions.
According to our previous analysis, the dynamics at $\kappa=0$ can be reasonably well described by including only the FGS $\Phi_0$ and one of the higher modes, say $\Phi_e$.
Consequently, we may obtain some intuition on the main features of the $\kappa\neq 0$ dynamics by considering the two corresponding QE bands: $\Phi_{0, \kappa}$ and $\Phi_{e, \kappa}$, so that 
the time evolution of $\Psi(x,0)$ reads 
(cf Eq. \ref{time}):
\begin{equation}
\begin{aligned}
 \Psi(x,mT) &= 
 \int_{-\pi / (n_p L)}^{+\pi / (n_p L)} d\kappa  \{ C_{0,\kappa} e^{-i\epsilon_{0,\kappa}mT/\hbar } \Phi_{0,\kappa}(x,0) \\
   &+ C_{e,\kappa} e^{-i\epsilon_{e,\kappa}mT/\hbar } \Phi_{e,\kappa}(x,0) \}
  \label{time2}
\end{aligned}
\end{equation}
Numerically, we find that the periodic part of the FGS $\Phi_{0,\kappa}$ and the associated QE $\epsilon_{0,\kappa}$ depend only very weakly on the quasi momentum $\kappa$, i.e. 
$\Phi_{0, \kappa} \approx e^{i\kappa x}\Phi_{0}$ and $\epsilon_{0,\kappa} \approx \epsilon_0$, which is in contrast to the excited state $\Phi_{e,\kappa}$ and $\epsilon_{e,\kappa}$
respectively. Thus, upon evaluating $|\Psi(x,mT)|^2$ we encounter a term proportional to $|\Phi_{0}(x,0)|^2$ plus terms including the $\kappa$ integral over $e^{-i\epsilon_{e,\kappa}mT/\hbar } \Phi_{e,\kappa}(x,0)$.
The latter ones, will quite generically inherit a diffusive decay due to the $\kappa-$dependence of $\epsilon_{e,\kappa}$ and thus tend to zero for large  
$mT$. Consequently we are left with $|\Psi(x,mT)|^2 \propto |\Phi_{0}(x,0)|^2$ and 
the occupation of the lattice sites becomes indeed time-independent and simply follows the lattice site occupations as given by the FGS $\Phi_{0}(x,0)$ which features
a strong desymmetrization between the lattice sites. Note that this simplified line of arguments neglects 
-among others- the diffusion caused by the 
weak $\kappa$-dependence of $\Phi_{0, \kappa}$ or by driving induced couplings of $\Phi_{0, \kappa}$ to higher bands which lead to the overall decay even at late times as seen in Fig.\ref{fig4} (b).

\paragraph*{Conclusion}
\label{S7}

Our study of the temporal evolution of the lattice site occupations for a quantum particle in a site-dependently driven lattice demonstrates
that we can achieve an almost complete depletion of certain lattice sites by exploiting narrow avoided crossings of the Floquet ground state with higher Floquet bands.
This represents a mechanism for a controlled site-selective loading of a lattice potential, 
which might open the roadway towards the 
preparation of initial states which are otherwise hard to obtain. 
The latter are certainly of interest for e.g. the investigation of the highly correlated many particle systems in optical lattices and their follow-up nonequilibrium dynamics.
In ultracold atomic physics Feshbach resonance management could be used to appropriately tune the interaction strength among the atoms \cite{Chin:2010} after 
the preparation of the corresponding initial state.
Experimentally, our results are of relevance whenever the translational symmetry over at least two neighbouring lattice sites is broken. In particular, for cold atom 
experiments performed in shaken optical lattices, there are several ways how this could be implemented, among them Fourier synthesized lattices where higher order Raman transitions allow 
for a substructure below half the lasers wavelength \cite{Weitz:2009}, distorted 2D lattices \cite{Hemmerich:2014} 
or so called 'painted potentials' where full control over the potential landscape is achieved \cite{Boshier:2009}.


\end{document}